\documentclass[11pt,letterpaper]{article}

\usepackage[paperwidth=8.5in,left=1.in,right=1.0in,top=1.0in,bottom=1.0in,
  paperheight=11.0in]{geometry}

\usepackage[detect-all]{siunitx} 
\usepackage{enumitem}
\usepackage{fancyhdr}
\usepackage[table]{xcolor}
\usepackage{bm}
\usepackage{graphicx}
\usepackage{newclude}
\usepackage{appendix}
\usepackage{amsmath,amsfonts}
\usepackage{hyperref}
\usepackage{wrapfig}
\usepackage[utf8]{inputenc}
\usepackage{pdfpages}
\usepackage{isotope}
\usepackage{geometry}
\usepackage{booktabs}
\usepackage{multirow}
\usepackage{bigstrut}
\usepackage{ragged2e}
\usepackage{units}

\usepackage{float}

\newif\ifsummary
\summaryfalse

\ifsummary 
\title {Snowmass Summary Report for AF5: Accelerators for Rare Processes and Physics Beyond
Colliders, Executive Summary}

\else
\title {Accelerators for Rare Processes and Physics Beyond
Colliders\\
Report of the AF5 Topical Group to Snowmass 2021}

\fi

\author{{\bf AF5 Convenors:}\\E. Prebys, UC Davis; Richard Milner, MIT; Mike Lamont, CERN\\
\\
{\bf Community Contributors:}\\
R. Bernstein, P. Huhr, D.Neuffer, F. Pellimone, E. Pozdeyev, V. Pronskikh, M. Toups,\\
R. Zwaska, FNAL; 
C. Barbier, ORNL; M. Calviani, CERN; T. Browder, U of Hawaii;\\ 
Tor Raubenheimer, SLAC; Y. Semertzidis, Korean Institute for Basic Science}
\begin{document}

\vskip .5 cm

\maketitle

\vskip 1 cm

\ifsummary

\else

\newpage

\tableofcontents

\section{Executive Summary}
\fi

\subsection{Overview}

A number of well-articulated fundamental physics questions invoke the need for exploration of rare processes and what has come to be called Physics Beyond Colliders (PBC). 
These initiatives are similar in spirit to those addressed by high-energy colliders, but require different types of beams and experiments. 
Modifications of existing accelerator complexes and future dedicated scientific infrastructure should be  considered for the next two decades through projects complementary to main stream applications of existing facilities.
The main areas of interest include:

\subsection*{Low energy hidden sector searches}

These are motivated by the QCD axion as well as astrophysical hints, the low energy hidden sector is potentially accessible via number of sub-eV Axion/ALP search techniques. For example:
\begin{itemize}
    \item Helioscopes (e.g. BabyIAXO/IAXO)
    \item Haloscopes using resonant cavities (e.g. ADMX) or other methods (e.g. MADMAX)
    \item Light-shining-through-walls experiments (e.g. JURA, STAX)
\end{itemize}

Many of these initiatives can profit from ongoing advances in accelerator technology e.g. high field superconducting magnets, superconducting RF, and are/could be considered at laboratories which host this technology and associated technical infrastructure.

\subsection*{Light Dark Matter searches}

Light Dark Matter searches in the MeV -- GeV mass range target a parameter space of the Hidden Sector of special relevance to open questions in cosmology. Options include:

\begin{itemize}
    \item  Direct detection WIMP searches (primarily addressed by the Cosmic Frontier).
    \item  Proton beam dump experiment: new proposals (e.g. BDF/SHiP), re-purposed existing experiments (e.g. NA62, MiniBooNE, SeaQuest)
    \item  Electron beam dump experiments: NA64, LDMX, BDX… Proposals include use of novel use of existing facilities (LCLS-II, CEBAF, FAST/IOTA, TRIUMF/ARIEL) or the development of new facilities.
    \item  Long lived particles at colliders (LHC, SuperKEKB)
\end{itemize}

\subsection*{Precision measurements and rare decays}

Precision measurements and rare decays can probe higher masses than accessible with direct searches, via searches for the possible influence of the contribution of loop diagrams in a number of scenarios. For example:

\begin{itemize}
    \item Ultra-rare or forbidden decays/reactions:
\begin{itemize}
    \item Kaon sector (NA62, KOTO, KLEVER)
    \item Lepton sector (TauFV, Mu3e, MEG,mu2e/mu2e-II)
\end{itemize}
     
   \item Precision measurements:
\begin{itemize}
    \item Permanent EDM: in protons/deuterons (CPEDM) or in strange/charmed baryons (LHC-FT)
    \item Anomalous magnetic moment of muon (g-2)
\end{itemize}
     
\end{itemize}

\subsection*{Technology}

Use of existing accelerator technologies or accelerator technology under development for novel physics applications in generally the PBC domain. 
Applications might include various types of axion/ALP searches (mentioned above), vacuum magnetic birefringence (VMB), exploration of Ultra-Light Dark Matter and Mid-Frequency Gravitational Waves (e.g. AION, MAGIS).
Technologies being exploited include:

\begin{itemize}
    \item High field magnets e.g. axion and dark matter searches
    \item Superconducting RF e.g. thin film SRF cavities for axion searches
    \item Induction LINACs e.g. rare muon processes
    \item Quantum sensors, ultra-sensitive opto-mechanical force sensors (e.g. KWISP), SQUID based measurement devices
    \item Carbon Nanotubes (CNT) (e.g. directional detection of DM candidates)
    \item Targetry, FFAs
\end{itemize}

\subsection{Organization of Report}

The mandate of the AF5 Working Group was to investigate ``Accelerators for Rare Processes and Physics Beyond Colliders" as outlined above.  This turned out to be something of an illusive topic, as in most cases, the accelerator needs of such experiments were already being investigated by their proponents.  
There was also significant overlap with other working groups within the 
Accelerator Frontier, specifically AF2 (Accelerators for Neutrinos) and AF7-T (Accelerator Technology - Targets and Sources).
There are also clear links to the work of the Rare Processes and Precision Measurement Frontier.

We elected not to investigate experiments that are entirely parasitic to other programs, such as neutrino beam dump experiments and high $\eta$ detectors at colliders, as we felt that those experiments themselves were not driving the accelerator choices.

In the end, we chose to divide our report as follows:
\begin{itemize}
    \item \textbf{Optimum exploitation of the PIP-II Linac at Fermilab}  The PIP-II Linac is a superconducting linear
    accelerator that will increase the injection energy of the Fermilab Booster from 400 MeV to 800 MeV.  
    The primary motivation for this is to increase the total proton power to the high energy neutrino program,
    specifically the DU$\nu$E Experiment;  however, that will only use about 1\% of the beam available
    from linac, which can provide a maximum of 1.6 MW at 800 MeV.
    
    Individual experiments, like Mu2e-II, have come forward with detailed proposals for the 800 MeV PIP-II beam, and other experiments but as yet there is no coherent plan for beam distribution and beam usage.  Other muon experiments like $\mu\rightarrow e\gamma$and $\mu\rightarrow 3e$ could also be carried out at this facility, although those proposals are not yet as developed.  Although it is not high energy physics, a surface muon program
    has been proposed as well.  
    
    To increase the sensitivity of the search for muon to electron conversion, an
    FFA has been suggested to produce pure muon beam with a narrower energy and time
    spread.  This would require a new bunch compressor ring to accumulate beam from
    the PIP-II linac and extract it in intense bunches at rates rates of 100-1000 Hz.
    Such a compressor could also serve a more extensive experimental program.  In particular,
    it could drive a set of dark sector searches based on beam dumps.
    
    We feel that the because of the time of the current Snowmass process relative
    to state of construction and planning for the PIP-II project, it can and should
    take a leading role in developing the experimental program.

     \item \textbf{Beam dumps} support a broad range of experiments, generally searching for dark sector or other
     rare particles.  These generally don't have demanding beam requirements beyond the total power and
     our role in this area will be to catalog the capabilities of various facilities, as these experiments use
     a wide range of beam energies
     
     Proton beam dump experiments range from energies in the GeV range, such as the RedTop Experiment at the CERN PS
     to the SHiP and NA62++ experiments, which use the 400 GeV beam from the CERN SPS.
     
     Electron beam dump experiments range from the 100 MeV beam for the DarkLight Experiment at JLab/TRIUMF to the 11 GeV
     BDX experiment at the same lab. Because electron beam dump experiments have no beam quality requirements
     beyond total beam power, they have potential as an early application of proton or laser wakefield acceleration,
     and charting a course for this line of research is another topic that should be discussed at Snowmass.

    \item \textbf{Other opportunities}  In this section, we will summarize diverse accelerator physics issues 
    related to specific fields of experimental goals, which are not automatically addressed in the quest
    for high energy or high intensity.  Of particular interest in this area is the Belle-II experiment at KEK,
    which has recently broken luminosity records.
    
    Storage rings for electric dipole moment (EDM) measurements also involve very interesting physics problems,
    particularly electrostatic storage rings of the sort that have been proposed to measure the
    proton EDM.

    \item Finally, we will discuss \textbf{synergistic activities with other R\&D areas}.  This section will highlight areas of 
    technological development that are primarily being developed for other areas of high energy physics which
    can also benefit BSM searches.  These include overlaps in target development with ongoing R\&D for neutrino
    or muon experiments, powerful magnet development for ``light through walls" experiments, and high Q RF development
    which in this case is primarily for axion searches.
    
\end{itemize}

\subsection{Recommendations}

These are interesting times for the field with well-articulated motivation for a multi-pronged exploration of unconstrained BSM parameter space.
A wide variety of possibilities have been proposed  under the Physics Beyond Collider banner. 
These span an impressive range of mass scales from low mass Axions and ALPS through to rare processes and EDM measurements that potentially probe the multi-TeV domain.

Exploitation of some of these options is an indispensable compliment to the main accelerator driven programmes.
Their competitiveness should be systematically evaluated along with the compatibility of their demands with already established, or planned, initiatives. 
An appropriate level of commitment should be planned in to allow full exploitation of the facilities and technology concerned.

We make the following recommendations:
\begin{enumerate}
    \item Given the high cost and long time scales of flagship energy frontier and 
    intensity frontier experiments, it is essential to support a broad range of
    smaller, near-term experiments, both to provide guidance for the larger experiments,
    and to provide training and experience to the next generation of physicists.
    \item The nature of dark matter is arguably the most important question in particle
    physics, so priority should be given to the wide variety of dark sector searches,
    particularly those involving electron or proton beam dumps.  To this end, an
    effort should be made to 
    exploit the facilities with the most potent capabilities for constraining the elusive dark sector. 
    This will be an ongoing process. The competitiveness of proposals in the worldwide landscape should be 
    evaluated, and 
    a critical evaluation of the demands from a facility perspective for each proposal e.g. required beam time, beam on target, cost of any upgrades or modifications should be performed.
    
    \item The current PIP-II project and the later Booster Replacement project at Fermilab are primarily driven by the needs of the DU$\nu$E experiment; however, there is a unique opportunity presented by taking full advantage of the excess protons to support a diverse low energy program, possibly including muon collider R\&D. 
    The potential, in terms of beam energy and protons-on-target should be quantified.
    It is important that a community of potential users be encouraged to evaluate the options to exploit this potential.  
    In particular, the specific choices made for the booster replacement should consider other scientific opportunities,
    rather than strictly focusing on eventually providing 2.4 MW to the LBNF.
\end{enumerate}

\ifsummary
\else

\section{Effective Exploitation of PIP-II and Beyond}

\label{sec:PIP-II}

\subsection{PIP-II Motivation and Scope}

While the LHC and its planned luminosity upgrades will be the flagship of the Energy Frontier for the foreseeable future, the US program based program is focusing on the
DU$\nu$E Experiment at the Sanford Underground Research Lab (SURF) in South Dakota.

The beam for this experiment will be provided by the new Long Baseline Neutrino Facility (LBNF) beam line and neutrino production target and horn at Fermilab.  To achieve the ultimate
physics goals of DU$\nu$E, beam power from the Fermilab Main Injector will need to
increase from the present level of 700 kW to 1.2 MW and later to 2.4 MW.  The beam power is currently limited
by space charge effects when injecting the 400 MeV beam from the present linac to
the Fermilab Booster, which then accelerates it to 8 GeV.

A key step to increasing the beam power is the PIP-II project.  The centerpiece of this
project is an 800 MeV Linac which will replace the current 400 MeV injection into the
Booster.  Because space charge tune shift scales as $1/\beta\gamma^2$, this will allow much
more beam to be injected into the Booster, particularly since it will be combined with a 
much more sophisticated ``painting" scheme to better distribute the beam transversely 
during injection.  The Booster repetition rate will also be increase from 15 to 20 Hz, and
the accelerating RF system in the Main Injector will be upgraded.  The cumulative effect of these improvements will be to increase the beam power to the LBNF beam line to 1.2 MW; however, this will only require about 1\% of the potential output of the PIP-II linac, which
can provide up to 1.6 MW of beam at 800 MeV.

The failure to fully and effectively exploit the capabilities of the PIP-II would not only 
be a waste of a unique and valuable resource, but would also call into question the
motivation for building such a complex facility in the first place.

A number of groups have expressed interest in the PIP-II beam, and some have developed
fairly detailed proposals, but there has as yet been no coordinated effort to
develop a cohesive 800 MeV program.  We feel that this presents a particularly
important opportunity and responsibility to the current Snowmass process.

\subsection{PIP-II Specifications}

The PIP-II Linac will provide a 800 MeV proton beam with continuous wave (CW) capability, with beam power up to $\sim1.6$ MW (2 ma, 800 MeV beam) available for user experiments \cite{PIP2-CDR}. 

\begin{table}[!h]
\caption{Beam Parameters for the PIP-II Linac.  The third and fourth columns assume an RF splitter and give the beam available to the ``central'' and ``side'' users, respectively.}
\label{tab:PIP-II}
\centerline{\begin{tabular}{| l | c | c | c | l |}
\hline
& {\bf Linac } & {\bf Central} & {\bf Side } & \\
{\bf Parameter }& {\bf Output} & {\bf Line} & {\bf Lines} & {\bf Comment} \\
\hline\hline
 Energy [MeV] &\multicolumn{3}{|c|}{800}&\\
\hline
Bunch Length [ps] & \multicolumn{3}{|c|}{4}&\\
\hline
Max. Ave. $H^-$/Bunch & \multicolumn{3}{|c|}{$0.8\times 10^8$} & 2 mA\\
\hline
Peak $H^-$/Bunch & \multicolumn{3}{|c|}{$2.0\times 10^8$} & 5 mA\\
 \hline
Bunch Frequency [MHz] & 162.5 & 81.25 & 40.625 & Maximum\\
 \hline
Bunch Separation [ns] & 6.2 & 12.3 & 24.6 & Minimum\\
\hline 
 \end{tabular}}
 \end{table}

The key parameters of the PIP-II Linac are shown in Table~\ref{tab:PIP-II}. It will produce proton bunches at 800 MHz with a maximum frequency of 162.5 MHz.  A proposed cryogenic upgrade will allow average currents of up to 2 mA, with peak bunch intensities of up to $2,0\times 10^8$ protons (5 mA).  A key feature of the linac will be a chopper in the Low Energy Beam Transport section that will allow arbitrary bunch patterns to be produced.  

The high energy program will only use about 1\% of the available beam pulses, limited by the 20 Hz repetition rate of the upgraded Fermilab Booster.  Although a final decision has not been made regarding the distribution of the remaining beam, a leading concept involves a 40.625 MHz RF deflector to split the beam into three sub-lines, as shown in Figure~\ref{fig:3split}.  Individual lines are selected by populating RF bunches of the appropriate phase.  We are assuming that Mu2e-II will have access to the central (node) line, and will therefore be able to receive bunches at up to 81.25 MHz.

\begin{figure*} 
    \centering
    \includegraphics[width=6in]{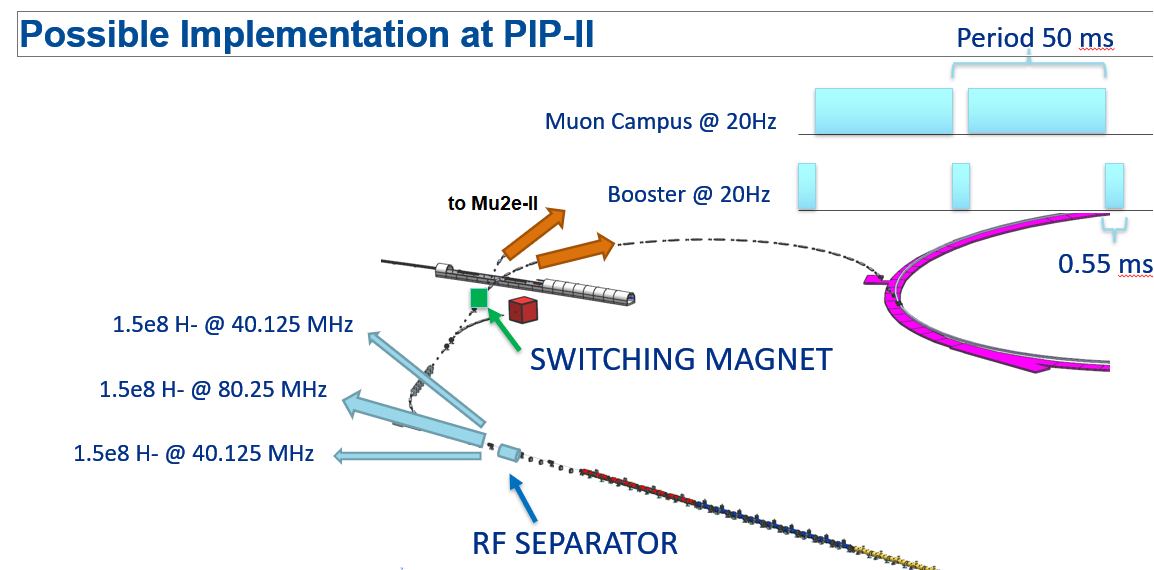}
    \caption{Layout of a possible implementation of multiple beam delivery at PIP-II. An rf separator is placed at the end of the Linac and a switching magnet separating beam for the booster from beam for other experiments are shown.}
    \label{fig:PIPIIsplit}
\end{figure*}

\begin{figure} [!h]
    \centering
    \includegraphics[width=4in]{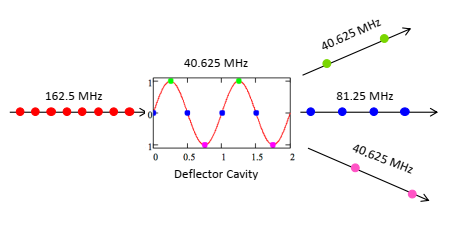}
    \caption{RF split of a 162.5 MHz beam into 3 beams by a 40.625 MHz deflector cavity. One of these beams would be for Mu2e-II, with the others directed toward other experiments.}
    \label{fig:3split}
\end{figure}

\subsection{Beyond PIP-II}

The PIP-II Project as described above is well defined and well along in the CD process. At the end of the project, the power to the 
LBNF line will increase to 1.2 MW.

To reach the 2.4 MW ultimately required by the DU$\nu$E experiment,
the still largely original Fermilab Booster will have to
be replaced.  

There are several options being considered for this.  They all
involve some increase of the energy of the PIP-II linac, with
some involving linear acceleration all the way to the Main
Injector injection energy while others will inject into an
intermediate rapid cycling synchrotron (RCS).  

In addition to beam power needs, some experiments require a bunch
structure that could not be served directly either by the output
of the PIP-II linac the output of an RCS.  For example and
upgraded muon to electron conversion experiment would utilize an
FFA, which requires a bunch structure of $\sim$10 ns bunches at
10-100 Hz.  

Such experiments would need some sort of accumulator ring, several
of which are currently at the conceptual stage.  

\subsection{Possible Experiments}

\subsubsection{Charged Lepton Flavor Violation (CLFV}

\begin{figure} [!h]
    \centering
    \includegraphics[width=4in]{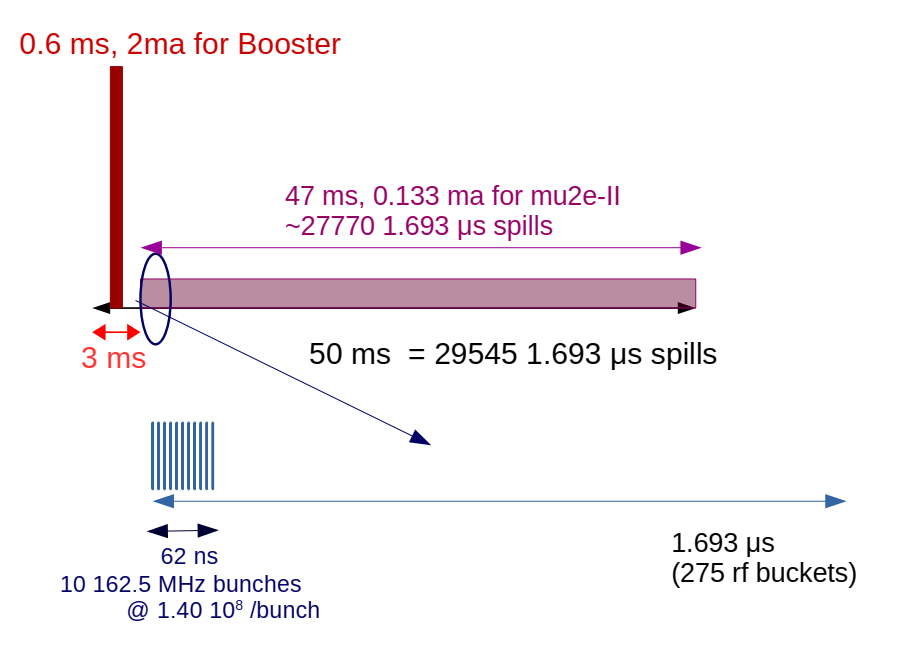}
    \caption{Formation of beam for Mu2e-II directly from the PIP-II Linac.}
    \label{fig:10bunch}
\end{figure}

The Mu2e-II Collaboration has developed a fairly detailed proposal to use
beam directly from the PIP-II linac\cite{MU2E-II-PIP2}.   The proposed bunch formation
is shown in Figure~\ref{fig:10bunch}.  During DU$\nu$E operation, PIP-II
bunch intensity is limited to $1.4\times 10^8$ by the 2 mA average beam current limit.
Ten such bunches every 1.6 $\mu$s would be about 100 kW of beam power, or about a factor of
10 more than the initial Mu2e experiment.  Once significant challenge will be injecting
the lower energy beam into the Mu2e Production Solenoid.

In addition to Mu2e-II, $\mu\rightarrow e\gamma$ and $\mu\rightarrow 3e$ experiments are also being considered.  Some of these could potentially use muons that had gone through the Mu2e experiment, although this might require deceleration by an induction linac or equivalent technology.

A full CLFV program based on PIP-II and potential upgrades is 
being delopeed and is discussed here\cite{NEW-CLFV}.

\subsection{Other Rare Decay Studies}

Rare decays have long played a role in providing hints for physics beyond the Standard Model,
and the mid-term program could provide a rich field for such experiments.

As mentioned previously, one of the most mature rare decay proposals for the PIP-II era is the REDTOP experiment, which would study the decay of $\eta$ and $\eta'$ mesons\cite{REDTOP}.  The initial proposal relies on the
Delivery Ring, so it assumes that Mu2e and g-2 are no longer using it.

The initial stage of the PIP-II Linac could drive a pion production experiment to
test lepton universality, as proposed by the PIONEER Collaboration\cite{LEPTON-UNIVERSALITY}.

There are a wide variety of studies that can be performed with kaons\cite{RARE-KAON-1,RARE-KAON-2,RARE-KAON-3}. Unfortunately, the 800 MeV PIP-II energy is too low to produce kaons. In addition, the linac portion
of any hybrid linac/RCS Booster replacement will likely also be below the kaon prodoction threshold.
A future kaon program would therefore have to be based on 8 GeV capacity in excess of 
what is required for the LBNF program.  If the Mu2e-II experiment ends up taking high power beam
from the 800 MeV PIP-II Linac, then there will be potentially up 100 kW of excess beam
at 8 GeV.  If it comes straight from the Booster, then it will come in 20 Hz strings of batches,
each about 1.6 $\mu$sec long.

It's possible that the requirements of some of these rare decay experiments will drive
the decisions that are made regarding such things as accumulator rings at intermediate
energies, so we reiterate our strong recommendation that the Fermilab low energy ($\le 8$ GeV)
community come together as soon as possible to coordinate with the plans for the
accelerator complex going forward.

\subsection{Muon Collider R\&D}

The high power, low energy proton beams at Fermilab will provide significant opportunities for 
Muon Collider R\&D.  This will be discussed in more detail in Section~\ref{sec:muon-collider}.
These opportunities should be discussed along with the other experimental programs.

\subsubsection{Slow Muon Program}

There has been discussion of a slow muon program based on the PIP-II linac\cite{SLOW-MU}, as
there is no such facility in the US.
The versatile bunch structure of the linac would allow it to deliver up to 3 kW of
beam in pulsed mode, making it competitive to the ISIS facility and up to 800 kW
in CW mode, making it competitive with world class cyclotrons like those at PSI and
TRUIMF.

\section{Beam Dumps}
\subsection{Introduction}

The lack of firm hints to the mass scale of new particles calls for a concerted effort by direct searches and precision measurements. 
At the same time the absence of new  particles is not necessarily due to their high scale of masses but could equivalently
be due to their weak scale of couplings with the Standard Model particles. 
This strongly motivates investing in a complementary programme to investigate the possibility of a light Hidden Sector coupled to the Standard Model.
Beam dump experiments offer one potential avenue for detection, direct or indirect, of such particles.

A limited number of facilities are capable of delivering protons at significant intensity and energy to deliver beam to proton beam dump experiments and all options are to be considered, if compatible with ongoing physics programs.
Electron beam dump experiments could be served by the more numerous electron machines. 
Muon beam dumps experiments are also an option at the limited number of world wide facilities.

The comparative physics reach for any proposal needs to be carefully evaluated, preferably against established benchmarks (see, for example the results of the  PBC BSM group's analysis \cite{PBC:BSM}). 
A realistic estimate of background, available beam time, PoTs or EoTs etc. needs to be fed in these evaluations.

The take home message is that beam dumps potentially offer significant physics reach in Hidden Sector searches. 
After evaluation and arbitration, they should be given some priority as part of a diverse programme.

\subsection{Proton Beam Dump Experiments}

In proton beam dump experiments, high fluxes of protons on a appropriate target could produce the Hidden Sector particles in meson decays.
Given the feeble coupling of the Hidden Sector particles, they would be (very) long-lived, and possibly decay back to Standard Model products within a long decay volume, up-steam of a suitable detector.
Beam dump in this context implies a target which aims at provoking hard interactions of all of the incident protons
and the containment of most of the associated cascade.  

Proton beam dump experiments are potentially superior to collider experiments in the 
sensitivity to GeV-scale hidden particles given potential luminosities several 
orders of magnitude larger than those at colliders. The large forward boost for light 
states, gives good acceptance despite the smaller angular coverage and allows 
efficient use of filters against background between the target and the detector, 
making the beam dump configuration ideal to search for new particles with long 
lifetimes.

Some already existing experiments (e.g. NA62-BD, MiniBooNE-DM, SeaQuest-BD) have repurposed to operate in a so-called beam-dump mode to performed dedicated dark sector searches.

\subsubsection{Summary of proposals and existing experiments}

Given the above motivation, there is already considerable activity as summarized in Table \ref{tab:pBDsummary} below. 

\begin{table}[htb]
\begin{center}
\caption{Brief overview of proposed and already realized proton beam experiments}
\label{tab:pBDsummary}
\begin{tabular}{lllcc}
\hline
Proposal &    Main Physics Cases  & Beam Line  & Beam Type \\
\hline

\textbf{MeV-GeV mass range:}  &  &  &  & \\
\hline
SHiP  & ALPs, Dark Photons, Dark Scalars &  BDF, SPS  & 400 GeV $p$   \\
 & LDM, HNLs, lepto-phobic DM, .. &  &  & \\
 
NA62 BD  & ALPs, Dark Photons,  & K12, SPS  & 400 GeV $p$  \\
 & Dark Scalars, HNLs  &  &  &  \\

RedTop  & Dark Photon, Dark scalar, ALPs  & CERN PS  & 1.8 or 3.5 GeV $p$  \\

SEAQUEST BD & LDM & FNAL MI & 120 GeV $p$ SX  \\ 
MiniBooNE-DM & LDM & FNAL Booster & 8 GeV/c $p$  \\

LANSCE-PSR Short-Pulse  & Dark Matter and Sterile Neutrino  & Los Alamos &   proton  \\

O(1 GeV) Proton  FNAL & Dark sector & Fermilab  & 1 GeV p  \\
O(10 GeV) Proton  FNAL & Dark sector  & Fermilab  & 10 GeV p  \\

\hline
\end{tabular}
\end{center}
\end{table}

At \textbf{CERN SPS} new Fixed Target facilities based on the SPS are considered with the combination of the high-intensity 
proton beam and slow beam extraction opening the way for a number of experiments in the \textbf{MeV -- GeV} range.
The design of a proposed Beam Dump Facility has been advanced to the conceptual design stage \cite{PBC:BDF}.
In the first instance, exploitation of the facility is envisaged to be for the Search for Hidden Particles (SHiP) experiment.
The TauFV experiment, which would operate in tandem with SHiP, is proposed to be installed in on the BDF beamline before the target. 
The aim is to intercept around 2\% of the incoming protons to look for ultra-rare $\tau$ decays.

Recently, the existing experimental cavern ECN3, at present used by NA62, has been considered a potential location for redesigned 
version of SHiP, or 
a new beam-dump experiment, SHADOWS, to search for a large variety of feebly-interacting particles possibly produced in the interactions of a 400~GeV proton beam with a high-Z material dump \cite{SHADOWS}.
SHADOWS will use the 400~GeV primary proton beam extracted from the CERN SPS currently serving the NA62 experiment in the CERN North area and will take data off-axis when the P42 beam line is operated in beam-dump mode.

At FNAL, \textbf{MiniBooNE-DM} has operated in beam dump mode where protons from the Fermilab Booster were delivered directly
to the steel beam dump of the Booster Neutrino Beamline (BNB). The MiniBooNE detector was used to search for the production of sub-GeV dark matter particles via vector-boson mediators.

\textbf{Seaquest-BD} performed early parasitic data taking 2019-2021 (POT ~\num{1.4e18}).
Phase-II from 2021 onwards foresees a possible detector upgrade, adding electron, photon and hadron capability
and a new dedicated dark matter programme.

Elsewhere, \textbf{LANSCE}, the Lujan neutron scattering center at the Los Alamos Neutron Science
Center, consists of a 800-MeV, short-pulse, 100-kW proton source and spallation neutron
source where such searches are ongoing with the Coherent CAPTAIN Mills (CCM) 10-ton, liquid argon
detector. The employment of fast timing coincidence of the beam with the detector is used to identify
signals and reject background. The current beam spill time width is 300 ns with intensity of \num{2.9e13}
protons per pulse at 20 Hz. With upgrades to the Proton Storage Ring (PSR), the beam spill width
may be compressed to 30 ns with minimal intensity loss enabling an increase in the signal to background
(S/B) of more than 100 and resulting sensitivity increase of an order of magnitude for dark matter and
sterile neutrino searches. This can be achieved with PSR accelerator upgrades on the time scale of a few
years and at modest cost. [Snowmass LoI 215]

\textbf{J-PARC} has studied the feasibility of an experiment searching for sub-millicharged particles using 30 GeV proton incident on a fixed-target.

\textbf{REDTOP} is a low-energy, fixed-target experiment\cite{REDTOP}. The proposal foresees
using a 1.8 GeV Continuous Wave (CW) proton beam impinging on a target made with 10 foils of a low-Z material (lithium or beryllium) to produce about \num{1e13} $\eta$ mesons in one year of running.
The physics motivation includes direct searches for dark photons and indirect searches for dark photons and new gauge bosons.
Realization at both CERN's PS and Fermilab has been explored.

\subsection{Electron Beam Dumps}

In an electron beam dump, high-intensity multi-GeV electrons impinging on the dump produce, potentially, dark matter candidates. 
Typically direct dark photon searches focus on identifying the mediator through its decay into SM particles. 
In the basic setup, a detector is placed downstream of the dump in location that minimises its exposure to background muons etc.

Other experiments combine an active beam dump and missing energy/momentum techniques. Here a thin target integrates with a number of sub-detectors.
The dark matter candidate may be produced via electron-nucleon scattering and identified through the missing energy/momentum carried away by the unseen particles. 

Again, there is already considerable activity in this domain, a selection of which are summarized in Table \ref{tab:eBDsummary} below. 

\begin{table}[htb]
\begin{center}
\caption{) }
\label{tab:eBDsummary}
\begin{tabular}{lllc}
\hline
Proposal &    Main Physics Cases  & Beam Line  & Beam Type  \\
\hline


LDMX & LDM & LCLS-II & 4/8 Gev  $e$  \\
BDX & LDM, A' & JLAB  & 11 GeV $e$  \\
DarkLight &	 A' & JLAB, TRIUMF  & 30-100 MeV $e$ on $A$ \\
SENSEI & LDM   & CCDs (FNAL/SNOLAB)  & cosmos  \\
MAGIX & A' & MESA  & 150 MeV $e$  \\
MMAPS &  $e^\text{-}e^\text{+}\rightarrow \gamma A'$ & Cornell synchrotron & 5.3 GeV $e^\text{+}$ SX     \\	
BELLE-II & ALPS, A' & SuperKEKB  & $e^\text{-}$$e^\text{+}$ $\sqrt{s}$ = 10.58 GeV  \\


\hline
\end{tabular}
\end{center}
\end{table}

The main US laboratory with a long term high intensity electron beam program for particle physics is JLab. Based on a superconducting linac, 
the beam was recently upgraded up to 12 GeV (JLab12) and can provide highly polarized e-pulses with unprecedented luminosity.
Besides the main programme, the high intensity beams also allows for searches of dark photons within the ongoing HPS and future APEX experiments.

TRIUMF has an electron-linac at the Advanced Rare IsotopE Laboratory (ARIEL).  At present, the maximum energy is 
30 MeV with intensities with up to 100 $\mu$A.   Experiment S2134~\cite{S2134} proposed by a Canadian-US collaboration has been approved~\cite{Cline2022} and is under construction to search for new physics in e+e- final-states.  An energy increase to about 60 MeV is under consideration~\cite{ARIEL-work}.

At SLAC extraction of a fraction of the LCLS-II CW 4 GeV beam (8 GeV in the future) to a facility primarily devoted to hidden sector searches. 
The LCLS-II/LCLS-II-HE is a 4 to 8 GeV SRF linac operating in a CW mode that is built to deliver bunches to a Free Electron laser (FEL) at a 1 MHz repetition rate.
LESA will use a high-rate kicker to extract a low-current, high-repetition-rate beam in between the FEL bunches to End Station A, a large experimental hall built in the
1960’s at SLAC. The low-current LESA beam can be either the dark current from the gun or sourced by a gun laser at 46 MHz or subharmonics thereof (with potential upgrades up to 186 MHz). 
This will provide a near-CW beam of electrons with currents ranging from pA to nA at energies of up to 8 GeV.
The headline experiment aiming exploit this potential is the \textbf{Light Dark Matter eXperiment (LDMX)}, an electron-beam fixed-target missing-momentum experiment that has unique sensitivity to light DM in the sub-GeV range.

CERN's \textbf{SPS}, which once accelerated electrons up to 22 GeV for injection into LEP, offers CERN an opportunity to create a new high intensity primary electron beam facility.
The eSPS collaboration have developed a design which foresees a Linac injecting 3.5 GeV electrons into the SPS. 
The electrons are then accelerated to 16 GeV and then slowly extracted to a new experimental facility. 
One potential application, proposed by the LDMX experiment, is a search for hidden particles in the MeV-GeV range with the missing momentum beam dump method.
\textbf{NA64} is an operational missing energy active electron beam dump experiment at the SPS North Area that foresees continued operation with electrons before moving to muons.

\textbf{Plasma wakefield accelerators} 
could have the potential to deliver high-energy electron and positron beams from 
compact accelerators.
Current research on plasma wakefield acceleration (PWFA) is focused on producing high-quality beams
in experiments operating at 10 Hz or less, but typical applications require beam rates in the 100 Hz - 100 MHz range. 
Dark Sector beam-dump-based searches require high-intensity, medium-energy beam drivers. 
Plasma wakefield acceleration has the potential to extend the energy reach of these experiments.
Beam-dump experiments as an application of a plasma wakefield accelerator based on beam parameters that the field aims to
demonstrate in the next decade, namely: energies in the range of 10 to 100 GeV, bunch intensities in the range of \num{1e8} to \num{1e10}
with beam emittances in the range of 1 to 100 $\mu$, and repetition rates up to 1 MHz. [Snowmass LoI 170].

As a PWFA exemplar, the proton driven \textbf{AWAKE facility} at CERN could be extended for beam-dump like experiments;
and the possibilities are elucidated in [Snowmass LoI 133].

Scientific opportunities for electron Dark Sector/WIMP searches and astrophysics measurements at the CBETA facility have also been proposed, more details are presented in the Synergies section of this document.

The beam dump/PBC  landscape is well motivated and fast moving. A full overview is not attempted here.
An overview of beam facilities worldwide is shown in table \ref{tab:facww} -- the PBC potential is significant.
As noted above, the competitiveness of any proposal in terms of physics reach and cost need to be carefully evaluated from a global perspective.

\newpage

\begin{table}[!htb]
\begin{center}
\caption{Beam facilities worldwide (* indicates study)}
\label{tab:facww}
\begin{tabular}{l l lll}
\hline\hline
\textbf{Lab}  & \textbf{Facility}           & \textbf{Beam}  & \textbf{Energy} & \textbf{Current/POT} \\
\hline
Cornell & Synchrotron & e+ & 5.3 GeV & $\sim$2.3 nA\\
&cBeta (ERL)& e- & 90 -- 270 MeV & 100 uA extr. \\
\hline
Fermilab & FAST & e- & 150 MeV &     \\
 & & protons & 2.5 MeV  &   \\
 & MTEST &  protons & 120 GeV  & \num{5e5}  \\
&&pi&1 -- 66 GeV  & \num{7e4} -- \num{1e6} \\
 & MCenter & p,pi &  $\geq$ 200 MeV  &   \\
& Booster & p for $\mu$  &  8 GeV &  \num{16e12} POT per cycle  \\
&  Booster      & p for $\nu$ &  8 GeV  &    \num{4.3e12} ppp  \\
& Main Injector & p for $\nu$ &  120 GeV  &   \num{49e12} ppp \\
\hline

SLAC   &  ESTB   &   e-    &    2-15 GeV  &  1 to 10$^9$ e-/pulse \\
        & LCLS II Sector 30*   &   e-  &    4 GeV  & low current, quasi-CW  \\
\hline

BNL & Booster  &  p  & 1.9 GeV  & \num{1.7e13}/bunch \\
 & AGS  &  p  & 24 GeV  & \num{1.2e13}/bunch \\
 & RHIC  &  p  & 24 -- 255 GeV  & \num{1.7e11}/bunch \\

\hline

TRIUMF &BL1A & p & 180-500 MeV  & 50 -- 75 kW  \\
       &ARIEL & e& 30-70 MeV & 5 kW \\
\hline
KEK & SuperKEKB & e+/e-  & 4 on 7 GeV   &  \num{8e35} cm$^{-2}$s$^{-1}$ \\
&PF photon factory & e- & 2.5 GeV &  450 mA \\
& PF-AR & e- & 6.5 GeV  & up to 60 mA  \\
& LINAC & e+/e-  & 10 GeV  &   \\
 & ATF & e- & 1.3 GeV  &  \num{1e10} e/bunch \\
\hline 
 J-PARC       &   LINAC &  H-  &   400 MeV  &  $\sim$45 mA \\
    &  RCS MLF  & p for n & 3 GeV & $\sim$1 MW  \\
  &  RCS MLF & p for $\mu$  &  3 GeV &   \\
  &   MR  FX  &  p for $\nu$  (T2K) &     30 GeV  & 490 kW \num{2.5e14} ppp  \\
  &   MR  SX  &  p for hadrons  &     30 GeV  &  51 kW \\
  &  MR &   p for $\mu$        &    8 GeV  &  3.2 kW (phase II:  56 kW)  \\

\hline
Novosibirsk &  VEPP-2000 &  e+e- collider   & 2 GeV CoM  &  \num{1e32} cm$^{-2}$s$^{-1}$ \\
            &   Super c-tau factory*  &e+e- collider   &   2 -- 5 GeV CoM  &  \num{1e35} cm$^{-2}$s$^{-1}$  \\

\hline        
Mainz  &  MAMI   &e-  &  1600 MeV  &  140 $\mu$A  \\
       &   MESA  &  e-   &  155 MeV  & 150 $\mu$A (1mA in ER mode)  \\
\hline       
JLab  & Hall-A  & e-  &  1-11 GeV  & 1 - 120 $\mu$A  \\
  & Hall-B  & e-  &  1-11 GeV  &  1 - 160 nA \\
  & Hall-C  & e-  &  1-11 GeV  &  2.5 - 150 $\mu$A \\
      & Hall-D  & e-  &  12 GeV   & 1-2 $\mu$A   \\
\hline      
Frascati &  Linac/BTF  & e+/e-  &  25 to 750 MeV   & 1 - 10$^{10}$ per pulse  \\
         &  DAFNE   & e+e- collider  &  1.02 GeV CoM  &   $\sim$1 A\\
\hline         
DESY   & Petra III & e-   & 6 GeV &  100 mA \\
& European XFEL & e- & 17.5 GeV  &  5 mA      \\
\hline
PSI   & Cyclotron      &   p for $\pi$ and $\mu$    &   590 MeV     &  2.4 mA (1.4 MW)      \\
  & Cyclotron      &   p for n  (SINQ)  &   590 MeV     &  2.4 mA (1.4 MW)      \\
  & Cyclotron      &   p for ultracold n   &   590 MeV     & 2-3\% duty factor full power   \\
  \hline\hline
\end{tabular}
\end{center}
\end{table}

\section{Other Physics Opportunities}

Although much of the high energy physics community is focused on the larger experiments
being discussed, there are numerous smaller experiments that have the potential
for high impact scientific results. 

Many of these parasitic or opportunistic in nature, so we will restrict ourselves to
proposals that involve challenging and interesting accelerator requirements.

\subsection{CP Violation in the B System (Belle-II)}

Since the termination of the BaBar experiment in 2008, SuperKEKB and Belle-II at KEK
have represented the flagship of the precision study of the B meson sector.  In addition
to the central goal of CP violation studies, the high luminosity and superb detector
resolution facilitate a broad array of precision physics measurements.

Both the accelerator and the detector have continued to improve over the years, and recently
SuperKEKB set a new luminosity record of $3.8\times 10^{34}~{\rm cm^{-2}s^{-1}}$, with
plans to ultimately reach $6.3\times 10^{35}~{\rm cm^{-2}s^{-1}}$. Two white papers were submitted to Snowmass regarding Belle-II plans.  The first was an estimate of beam backgrounds
at the ultimate luminosity~\cite{BELLE-II-BACKGROUNDS} and the other was from a working
group working on plans for a polarized electron beam~\cite{BELLe-II-POL}.  Among other things,
this would enable the measurement of the anomalous g-2 of the $\tau$ lepton, which would
be complementary to the ongoing $\mu$ g-2 measurements.

The upgrades of the SuperKEKB accelerator will also allow increasingly sensitive searches for possible new physics beyond the Standard
Model in flavor, tau, electroweak and dark sector physics that are both complementary to and competitive with the LHC as well as other experiments.

\subsection{Electric Dipole Moment (EDM) Measurements}

The study of electric dipole moments in any experimental system (electron, neutron,
proton, atom, molecule) is an established and complementary way to probe mass scales well beyond
the reach of colliders.  For example, the limits on the existence of the neutron EDM have monotonically decreased over decades and have been one of the most stringent eliminators of new physics. 
A new experimental approach is under active consideration, namely to use unique storage rings that
have very challenging specifications, which are not naturally addressed within
the more mainstream areas of accelerator physics research.  An excellent overview
of the field of EDM measurements was submitted as a white paper to Snowmass~\cite{EDM}.

The experimental requirements for charged particle EDM searches using storage rings are very demanding and require the development of a new class of high-precision, primarily electric storage rings~\cite{EDM-store}. Precise alignment, stability, field homogeneity, and shielding from perturbing magnetic fields play a crucial role. Beam intensities around $N=4 \times 10^{10}$ particles per fill with a polarization of $P=0.8$ are anticipated. Large electric fields of $E=10$ MV/m and long spin coherence times of about $\tau_{SCT} = 1000$ s are necessary.  Efficient polarimetry with large analyzing power of $A_y \approx 0.6$, and high efficiency of detection $f \approx 0.005$ need to be provided.  In terms of the above numbers, this would lead to statistical uncertainties of
$$
\sigma_{stat} = \frac{2 \bar{h}}{\sqrt{Nf} \tau_SCT P A_y E}
$$
yielding $\sigma_{stat}(1\ \rm year)= 1.9 \times 10^{-29}$ e$\cdot$cm, where for one year of data taking 10000 cycles of 1000 s duration is assumed. The experimentalist’s goal must be to provide systematic uncertainties to the same level.

The JEDI (J{\"u}lich Electric Dipole moment Investigations) collaboration is pursuing the project of a precision storage ring to measure the EDMs of polarized proton and deuteron beams with unprecedented sensitivity. Besides R\&D covering, e.g., prototype development and spin dynamics simulations, a precursor experiment with polarized deuterons has been conducted at the Cooler Synchrotron COSY of Forschungszentrum J{\"u}lich (Germany), and is currently being analyzed. The next step of the research will require a new class of hitherto not existing precision storage rings. This encompasses the complete chain from design and building to operation of the storage ring and includes instrumentation for control/feedback of the beam(s) and its polarization. Such a ring requires a large electric field ($\approx$ 10 MV/m) and a proton momentum of 707 MeV/c, resulting in a storage ring of $\approx$ 500 m circumference. Recently, it was realized that, before starting the construction of a such a ring, an intermediate step, a so-called prototype with approximately 100 m circumference, is needed, which demonstrates the functionality of all components and allows for a first direct measurement of the proton EDM as well as searches for axions/axion-like-particles~\cite{EDM-Kim}. The current state-of-the-art of the development is summarized in a recent CERN Yellow Report~\cite{CERN-Abu}. 

The U.S. has the option of putting the ring inside the 805 m AGS ring-tunnel at Brookhaven National Laboratory (BNL), saving the cost of the tunnel. This will result in a comfortable electric field of 4.4 MV/m, which does not require the time and expense of a prototype ring with a 10 MV/m electric field, beyond the present state of the art. 

\subsection{Axion helioscopes} 

IAXO is a large-scale axion helioscope that will look for axions and axion-like particles (ALPs) produced
in the Sun with unprecedented sensitivity. The near term goal of the collaboration is the construction
and operation of BabyIAXO, an intermediate experimental stage of IAXO that will be hosted at DESY.

BabyIAXO is conceived to test all IAXO subsystems (magnet, optics and detectors) at a relevant scale
for the final system and thus serve as prototype for IAXO, but at the same time as a fully-fledged
helioscope with relevant physics reach in itself, and with potential for discovery. BabyIAXO is now
under construction and its commissioning is expected for 2023. The BabyIAXO baseline program would
ideally overlap with the IAXO design and construction with a provisional commissioning in 2027.

\section{Synergistic Opportunities with Other Working Groups}

In addition to accelerators themselves, a number of dark sector and other rare process rely
on accelerator technology, but may have needs that differ from the more mainstream
HEP research sectors.  It's therefore important to open lines of communication 
to these areas and to discuss synergistic R\&D areas as part of the Snowmass process.

Within the current Snowmass framework, technology development falls under Accelerator
Frontier working group 7 (AF7), which is further divided into three subgroups.
\begin{itemize}
    \item AF7-T: Targets and Sources
    \item AF7-M: Magnets
    \item AF7-R: RF
\end{itemize}

The work of all of these groups is summarized at the Snowmass website\cite{AF-WG-REPORTS}.

There are also significant synergistic opportunities between the 
muons collider collaboration and a number of the rare muon studies, particularly
those discussed for PIP-II and beyond in Section~\ref{sec:PIP-II}.

In addition, high power lasers, which did not have a separate dedicated study group, have 
several applications in dark sector searches.

\subsection{Targets and Sources}

Work in this area is primarily focused on high power targets for the neutrino and 
possibly muon sectors.  To this end, there are ongoing investigations of 
novel materials and target designs as well as extremely radiation hard instrumentation
for targets.

All of this work could benefit beam dump experiments, particular the latter in radiation
hard instrumentation, in that it instrumented smart targets could be useful to
a number of dark sector searches.

\subsection{Magnets}

The most significant driving force behind magnet development in High Energy Physics is the
desire to go to higher fields to keep the next generation of proton colliders to a manageable size.  The biggest single thrust in this area is the development
of magnets based on the Nb$_3$Sn superconductor, which could provide fields on the order of 15 T, compared to the approximately 8 T limit of traditional NbTi magnets.  An impressive amount of work has been done in this area over the last two decades.
Beyond Nb$_3$Sn, the community look toward hybrid magnets using high temperature superconductors that could conceivable go to 20 T or beyond; however, there is a long R\&D road ahead before such magnets
become practical.

Solar Axion searches, haloscopes, Axion and WISP searches using the "Light shining through walls" (LSW) approach all call on 
large aperture, high field magnets in which the high magnetic field couples an Axion or ALP to an observable photon.
Ongoing magnet R\&D, both general and specific, could clearly benefit this research. See \cite{Siemko:2652165} for a European perspective on the possibilities.

\textbf{Specific magnet R\&D for Low-Mass Axion Searches}
most axion dark matter searches take advantage of the coupling between axions and photons in the
presence of a magnetic field. Up until now, most experiments have used stock magnet designs to generate
this coupling. DMRadio and future axion searches have reached a scale where significant magnetic-field
engineering is required to enhance sensitivity of axion searches into the QCD band. These next generation experiments will require larger fields without sacrificing volume while also taking into account several
practical considerations, such as the ability to be cooled to cryogenic temperature, in order to reach these
sensitivity goals. Next-generation experiments will need to collaborate with national magnet labs, in addition to partnerships with industry, to design, construct and test the magnets that will power these upcoming
axion searches [Snowmass LoI 244].

The powerful future neutrino experiment facilities enable searches for Beyond the Standard Model phenomena in these experiments.
Of these, low-mass dark matter and other charge neutral particles could also be produced in the neutrino target and the beam dump. 
The concept for a \textbf{Neutral-Rich Three-Dimensional Sign-Selecting Focusing System} discusses an idea of utilizing a three dimensional
sign-selected focusing horn train system that would permit the co-existence of beam-dump style experiments
and the precision neutrino experiments. The key component in this system would be a three dimensional dipole
that would direct horn focused charged particle beams toward the direction of the neutrino experiments with minimal
loss [Snowmass LoI 209].

\subsection{RF}

The cutting edge of RF R\&D are high-Q high gradient $\pi$ cavities being developed 
for for linear accelerators.  This field leveraged a lot of R\&D for the 
high energy e$^+$e$-$ colliders, but so far it's application has been elsewhere.  For example, the proton linacs used at the SNS and ESS spallation neutron sources and
the PIP-II linac at Fermilab which was discussed previously. This is also the technology that's enabling the LCLS project at SLAC.

Several Dark Sector initiatives such as axion dark matter (DM) searches, such as the 
haloscope, helioscope proposals, require an extensive use of RF technologies.
For the resonant photon conversions of axions or ALPs in a strong background magnetic field, the
development of resonant cavities of various sizes and shapes with quality factor of \num{e5} to \num{e6} could profit
from the current RF technologies. To probe axion of mass in the range 0.1 to 800 eV, 
the frequency must cover the domain 0.2 to 200 GHz. To fully exploit the conversion volume permeated by
the strong magnetic field, multiple cavity schemes could be implemented at high frequency. This implies
the production of identical cavities and tuning all of them at the same frequency with dedicated phase
matching.
It is certainly important to keep open lines of communication between the two communities.

\subsection{Lasers}

State-of-the-art laser technology is regularly utilized in modern precision metrology.  Typically, sophisticated techniques are implemented to stabilize commercial lasers to very high precision.  This has significant overlap with the laser technology used by LIGO for gravitational wave detection.  Further, cryogenic techniques can be used to reduce noise~\cite{Kenn2020}. 

The lasers are used to cool atoms, for precision timing, to interrogate energy levels in atoms, and to manipulate quantum states of atoms.
This allows experiments to set limits on ultralight dark matter, typically by carrying out spectroscopy sensitive to time variation of both the fine structure constant and the electron mass~\cite{Kenn2020,NEW-HORIZONS-2022}.  This technology is also highly relevant for development of quantum computing.  Further, lasers are used for optical pumping to realize new sources of polarized ions~\cite{Max2020} for RHIC/EIC.

\subsection{Muon Collider}

\label{sec:muon-collider}
There have been significant developments in Muon Collider R\& D since the last Snowmass process,
as well as increased interest in Europe.  The Muon Collider Forum has
produced a very thorough report\cite{MUON-COLLIDER}.

There are clearly lots of opportunities for collaboration between the Muon Collider 
groups and groups interested in muon physics in general, and particular attention should be
paid to potential opportunities for PIP-II and beyond at Fermilab, as discussed in 
Section~\ref{sec:PIP-II}.

\section*{Appendix: AF5 Proposals}

For completeness, the full list of AF5 proposals is shown in Figure \ref{fig:AF5P}

\begin{figure} [!h]
    \centering
    \includegraphics[width=6in]{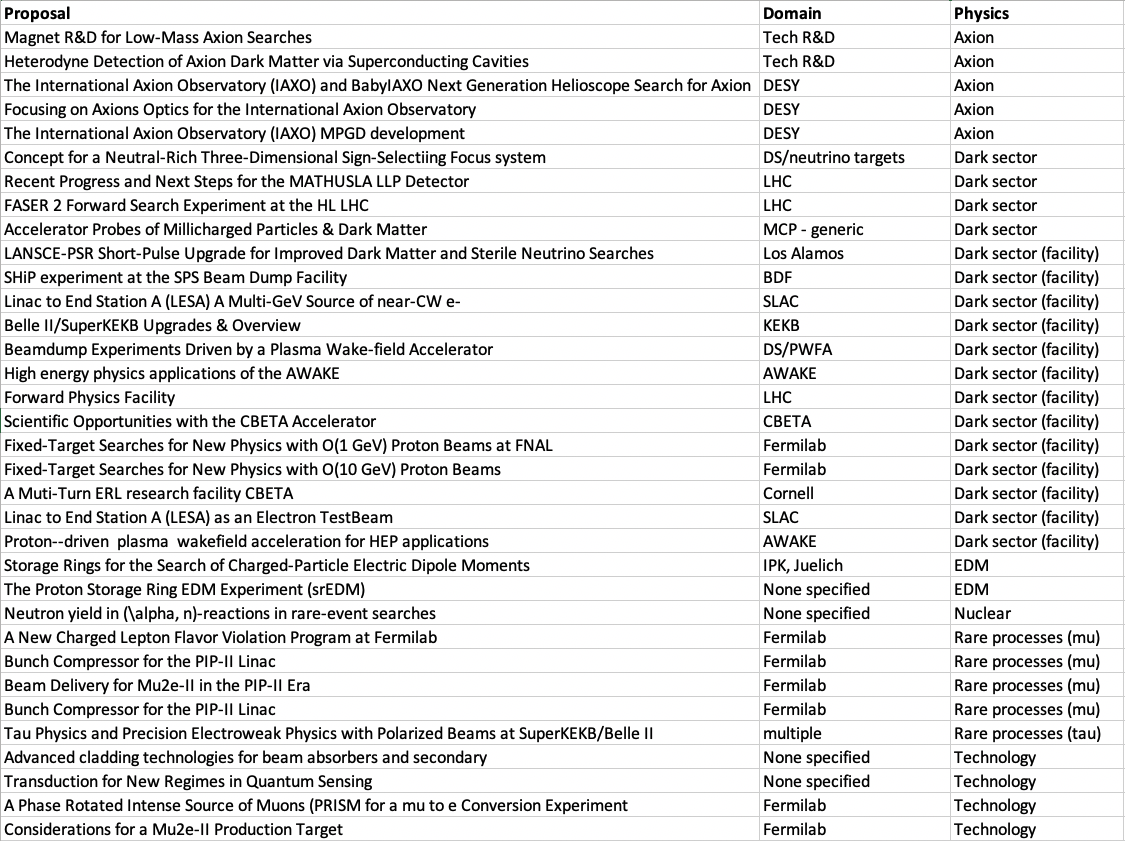}
    \caption{AF5 proposals}
    \label{fig:AF5P}
\end{figure}

\bibliographystyle{utcaps_mod}

\bibliography{bibliography/refs.bib}

\providecommand{\href}[2]{#2}\begingroup\raggedright\begin{thebibliography}{10}

\bibitem{PIP2-CDR}
M.~Ball {\em et~al.}, ``{\em The PIP-II Conceptual Design Report},''
  \href{http://dx.doi.org/10.2172/1346823}{ (3, 2017) }.
  \url{https://www.osti.gov/biblio/1346823}.

\bibitem{MU2E-II-PIP2}
K.~Byrum {\em et~al.}, ``{\em Mu2e-II: Muon to electron conversion with
  PIP-II},'' 2022.
\newblock \url{https://arxiv.org/abs/2203.07569}.

\bibitem{NEW-CLFV}
M.~Aoki {\em et~al.}, ``{\em A New Charged Lepton Flavor Violation Program at
  Fermilab},'' 2022.
\newblock \url{https://arxiv.org/abs/2203.08278}.

\bibitem{REDTOP}
C.~Gatto, ``{\em The REDTOP experiment},'' 2019.
\newblock \url{https://arxiv.org/abs/1910.08505}.

\bibitem{LEPTON-UNIVERSALITY}
{PIONEER Collaboration}, W.~Altmannshofer, {\em et~al.}, ``{\em Testing Lepton
  Flavor Universality and CKM Unitarity with Rare Pion Decays in the PIONEER
  experiment},'' 2022.
\newblock \url{https://arxiv.org/abs/2203.05505}.

\bibitem{RARE-KAON-1}
T.~Blum {\em et~al.}, ``{\em Discovering new physics in rare kaon decays},''
  2022.
\newblock \url{https://arxiv.org/abs/2203.10998}.

\bibitem{RARE-KAON-2}
J.~Aebischer, A.~J. Buras, and J.~Kumar, ``{\em On the Importance of Rare Kaon
  Decays: A Snowmass 2021 White Paper},'' 2022.
\newblock \url{https://arxiv.org/abs/2203.09524}.

\bibitem{RARE-KAON-3}
T.~KOTO, {LHCb}, N.~Collaborations, and t.~U. K.~I. Group, ``{\em Searches for
  new physics with high-intensity kaon beams},'' 2022.
\newblock \url{https://arxiv.org/abs/2204.13394}.

\bibitem{SLOW-MU}
E.~Prebys, A.~Hillier, and S.~Sheehy, ``{\em {Proposal to Use the Fermilab
  PIP-II Linac to Support a Low Energy Muon Program}},'' 2016.
\newblock
  \url{https://beamdocs.fnal.gov/AD/DocDB/0054/005489/001/pip-II-muons.pdf}.
  FERMILAB-BEAM-DOC-5489.

\bibitem{PBC:BSM}
{\bfseries BSM Working Group} Collaboration, P.-P.~B. WG, ``{\em {Physics
  Beyond Colliders: BSM Working Group Report}},''
  \href{http://arxiv.org/abs/1901.09966}{[{\ttfamily 1901.09966}]}.
  \url{http://cds.cern.ch/record/2652223}.

\bibitem{PBC:BDF}
P.-A.~B. WG, ``{\em {SPS Beam Dump Facility Comprehensive Design Study }},''.
  \url{https://cds.cern.ch/record/2650896}.

\bibitem{SHADOWS}
W.~Baldini, A.~Balla, J.~Bernhard, A.~Calcaterra, V.~Cafaro, N.~Charitonidis,
  A.~Ceccucci, V.~Cicero, P.~Ciambrone, H.~Danielsson, A.~De~Roeck, F.~Duval,
  G.~D’Alessandro, G.~Felici, L.~Foggetta, L.~Gatignon, A.~Gerbershagen,
  V.~Giordano, G.~Lanfranchi, I.~Lax, A.~Montanari, R.~Murphy, A.~Paoloni,
  G.~Papalino, T.~Rovelli, A.~Saputi, S.~Schuchmann, F.~Stummer, G.~Torromeo,
  N.~Tosi, and A.~Vannozzi, ``{\em {SHADOWS (Search for Hidden And Dark Objects
  With the SPS): Expression of Interest}},'' tech. rep., CERN, Geneva, Jan,
  2022.
\newblock \url{https://cds.cern.ch/record/2799412}.

\bibitem{S2134}
J.~Bernauer, R.~Corliss, R.~Milner, {\em et~al.}, ``{\em Search for New Physics
  in e$^+$e$^-$ Final States With an Invariant mass of 13-17 MeV using the
  ARIEL Electron Accelerator},'' April, 2021.
\newblock Approved with high priority by TRIUMF Particle Physics Experimental
  Evaluation Committee.

\bibitem{Cline2022}
E.~Cline and others (The DarkLight~Collaboration), ``{\em Searching for New
  Physics with DarkLight at the ARIEL Electron-Linac},'' 2022.
\newblock \url{https://arxiv.org/abs/2208.04120}.

\bibitem{ARIEL-work}
J.~Bernauer, R.~Corliss, M.~Hasinoff, R.~Kanungo, J.~Martin, R.~Milner,
  K.~Pachal, and S.~Yen, ``{\em TRIUMF Workshop on New Scientific Opportunities
  at the ARIEL electron linac, May 25-27, 2022},'' 2022.
\newblock \url{https://meetings.triumf.ca/event/262/timetable/#20220525}.

\bibitem{BELLE-II-BACKGROUNDS}
A.~Natochii {\em et~al.}, ``{\em Beam background expectations for Belle II at
  SuperKEKB},'' 2022.
\newblock \url{https://arxiv.org/abs/2203.05731}.

\bibitem{BELLe-II-POL}
T.~Markiewicz, T.~Raubenheimer, N.~Toro, and m.~o. t. L.~c. team, ``{\em The
  SLAC Linac to ESA (LESA) Beamline for Dark Sector Searches and Test Beams},''
  2022.
\newblock \url{https://arxiv.org/abs/2205.13215}.

\bibitem{EDM}
R.~Alarcon {\em et~al.}, ``{\em Electric dipole moments and the search for new
  physics},'' 2022.
\newblock \url{https://arxiv.org/abs/2203.08103}.

\bibitem{EDM-store}
F.~Rathmann and N.~N. Nikolaev, ``{\em Electric dipole moment and searches
  using storage rings},'' 2019.
\newblock \url{https://arxiv.org/abs/1904.13166}.

\bibitem{EDM-Kim}
O.~Kim and Y.~Semertzidis
  \href{http://dx.doi.org/10.48550/ARXIV.2203.08103}{Phys. Rev. D {\bfseries
  104} no.~1, (2021) 096006}.

\bibitem{CERN-Abu}
F.~Abusaif and others (CPEDM), ``{\em Storage Ring to Search for Electric
  Dipole Moments of Charged Particles - Feasability Study (CERN, Geneva
  2021)},'' 2019.
\newblock \url{https://arxiv.org/abs/1912.07881}.

\bibitem{AF-WG-REPORTS}
``{\em {AF Reports (Drafts)}}.''
\newblock \url{https://snowmass21.org/accelerator/start#af_reports_drafts}.

\bibitem{Siemko:2652165}
A.~Siemko, B.~Dobrich, G.~Cantatore, D.~Delikaris, L.~Mapelli, G.~Cavoto,
  P.~Pugnat, J.~Schaffran, P.~Spagnolo, H.~Ten~Kate, and G.~Zavattini, ``{\em
  {PBC technology subgroup report}},'' tech. rep., CERN, Geneva, Dec, 2018.
\newblock \url{https://cds.cern.ch/record/2652165}.

\bibitem{Kenn2020}
C.~J. Kennedy, E.~Oelker, J.~M. Robinson, T.~Bothwell, D.~Kedar, W.~R. Milner,
  G.~E. Marti, A.~Derevianko, and J.~Ye
  \href{http://dx.doi.org/10.1103/PhysRevLett.125.201302}{Phys. Rev. Lett
  {\bfseries 125} no.~20, (2020) 201302}.

\bibitem{NEW-HORIZONS-2022}
D.~Antypas {\em et~al.}, ``{\em Snowmass 2021 White Paper - New Horizons:
  Scalar and Vector Ultralight Dark Matter},'' 2022.
\newblock \url{https://arxiv.org/abs/2203.14915}.

\bibitem{Max2020}
J.~Maxwell {\em et~al.}
  \href{http://dx.doi.org/10.1103/PhysRevLett.125.201302}{Nucl. Instr. and
  Meth. A {\bfseries 959} no.~20, (2020) 161892}.

\bibitem{MUON-COLLIDER}
K.~Black {\em et~al.}, ``{\em {Muon Collider Forum Report (DRAFT)}}.''
\newblock
  \url{https://indico.fnal.gov/event/55041/attachments/156225/204232/MC_Forum_Report-v06232022.pdf}.

\end{thebibliography}\endgroup


\providecommand{\href}[2]{#2}\begingroup\raggedright\endgroup
\fi
\end{document}